\documentclass[aps,pra,preprint]{revtex4}
\usepackage{amsmath}
\usepackage{amsfonts}

\def\ra{\rangle}
\def\la{\langle}

\begin{document}

\title{Quantum Work Relations under trial Hamiltonians}

\author{Arun K.\ Pati$^{(1)}$\footnote{Email: akpati@iopb.res.in}, 
Mamata Sahoo$^{(1,2)}$ and Biswajit Pradhan$^{(1)}$}

\address{$^{(1)}$Institute of Physics, Bhubaneswar-751005, Orissa, India }

\address{$^{(2)}$ Max Planck Institute for Colloids and Interfaces, 
Science Park Golm, D-14424 Potsdam, Germany}


\date{\today}

\begin{abstract}
The universal quantum work relation connects a functional of an arbitrary 
observable averaged over the forward process to the free energy difference and 
another functional averaged over the 
time-reversed process. Here, we ask the question if the system is driven out of 
equilibrium by a different Hamiltonian rather than the original one 
during the forward process and similarly during the reversed process then how 
accurate is the quantum work relation. We present an inequality
that must be satisfied when the system is driven out by such a trial 
Hamiltonian. This also answers the issue of accuracy of the Jarzynski relation 
with a trial Hamiltonian. We have shown that the correction term can be 
expressed as the averages of the difference operator between the accurate 
and trial Hamiltonians. This leads to a generalized version of 
the Bogoliubov inequality for the free energy differences.
\end{abstract}

\maketitle

\section{Introduction}
In recent years work relations in nonequilibrium setting have attracted much 
attention. Jarzynski's equality \cite{jar} is an important step in that 
direction, where this provides a 
relation between the distribution of work performed on a classical system 
subjected to an external force and the free energy difference between the 
initial and final configurations. In particular, a finite system is prepared 
in a state of equilibrium with an environment at temperature T. The system is
driven with an external parameter from some initial value to  a final value.
During this process the Jarzynski equality connects the equilibrium information
about the free energy to that of the nonequilibrium processes.
The nonequilibrium work relation has been proved from a generalized version of
the fluctuation theorem \cite{crook}.
The Jarzynski equality has been generalized to quantum domain also 
\cite{kur,monnai,tasaki,tal}. Kurchan has
considered extension of this relation in a quantum setting where he has used 
measurement based schemes for the system of interest at the initial and final 
time and related the work performed to
the energy difference between the final and initial eigenstates \cite{kur}. 
Monnai  \cite{monnai} has derived the quantum version of the Jarzynski 
equality in terms of the microscopic reversibility and has explored the 
equilibrium information about the free energy of a system when it is thermally 
isolated and is driven externally out of equilibrium.
The derivation of the work relation by Monnai \cite{monnai} and Tasaki 
\cite{tasaki} is based on the twice measurements scheme. For more details 
regarding the twice measurements scheme, readers are advised to see 
\cite{and,mona}. In addition, there is quantum work operator approach to 
derive the work relation which can give the required energy
difference when the system is driven out of equilibrium \cite{montas}. 
The quantum analog of 
the Jarzynski equality 
has also been derived for closed \cite{kur,tasaki,muka} as well as open 
quantum systems \cite{hangi}.

Very recently, Andrieux and Gaspard \cite{gas} have proved a universal quantum 
work relation which connects the average of a functional of an arbitrary 
observable 
during the forward process to that of another functional averaged over the 
time-reversed process and the free energy difference. 
As shown in that paper many well known results follow
from this universal work relation. For example, the Jarzynski equality 
\cite{jar}, quantum 
Green-Kubo relations \cite{kubo}, and Casimir-Onsager \cite{on} 
relations can be obtained from the universal work relation.

In statistical mechanics approximate methods are very useful when dealing 
with complex systems that consist of many particles. 
For instance computing the free energy and the partition function of 
statistical systems are in general intractable. Only for few special models 
one may be able to calculate such thermodynamic quantities. In these situations
one replaces the original Hamiltonian (which may be difficult to handle) by 
a trial Hamiltonian (or an approximate Hamiltonian). Under such approximations, 
proving various thermodynamic relations are important and one must try to
establish how accurate are those relations.
In this paper, we ask the question if the system is driven out of 
equilibrium by a different Hamiltonian $H'(t)$ rather than the original one 
during the forward process and similarly during the reversed process then how 
accurate is the universal quantum work relation. 
For complex systems testing the  
Jarzynski equality or the Andrieux-Gaspard relation may be equally difficult. 
If we imagine that $H'(t)$ is a trial Hamiltonian then the inequality that we 
are going to prove in the sequel will 
tell us how accurate is the universal work relation or the Jarzynski relation
as compared to the actual one. In general, $H'(t)$ may contain different 
interaction or several external control parameters.
In these cases this inequality will be of use in estimating the work performed
on the system and the free energy differences.
In particular, we have shown that the correction term can be explicitly 
expressed as the averages of the difference operator between the 
accurate and trial Hamiltonians. Therefore, the error in the free energy 
difference caused by the modelling procedure can be estimated by calculating 
the difference operator. This in turn leads to a generalized version 
of the Bogoliubov inequality for the free energy differences.

 The organization of our paper is as follows. In section II, we introduce the 
universal work relation. In section III, we prove the main inequality for 
the universal work relation under the trial Hamiltonian. In section IV, we 
address the issue of accuracy of the Jarzynski relation under the trial 
Hamiltonian. Also, we prove a generalized version of the Bogoliubov inequality 
for the free energy differences. In section V, we present yet another useful 
inequality for the Jarzynski relation in terms of the operator norm. Finally, 
we conclude with some discussions in section VI.

\section{ Universal work relation}

Before presenting the main result let us briefly introduce the setting in which 
the universal work relation of Andrieux and Gaspard \cite{gas} was proved. 
Let us imagine a system whose Hamiltonian is given by a Hermitian 
operator $H(t,{\bf R})$, where  ${\bf R}$ can be some external parameters that 
change sign under the time-reversal
(such as the magnetic field). Under the time reversal operator $\Theta$ the 
Hamiltonian transforms as $\Theta H(t,{\bf R}) \Theta =  H(t,- {\bf R})$, with
$\Theta$ being an anti-linear operator and $\Theta^2 = I$. Initially 
the system is prepared in thermal equilibrium with the inverse temperature 
$\beta = \frac{1}{k_B {\rm T}}$, where $k_B$ being the Boltzmann constant and 
${\rm T}$ is the temperature.  During the forward process, the system is 
driven out of equilibrium by a time dependent Hamiltonian $H(t,R)$.
The initial state of the system is a canonical density matrix

\begin{eqnarray}
\rho(0) = \frac{e^{-\beta  H(0,{\bf R}) } }{Z(0)}, 
\end{eqnarray}
where $Z(0) =  {\rm Tr}[e^{-\beta  H(0,{\bf R})}]$ is the partition function, 
with $\rho(0)^{\dagger} = \rho(0), 
\rho(0) > 0$ and ${\rm Tr} \rho(0) =1$.
The free 
energy $F(0)$ of the system at the initial time is given via the 
partition function $Z(0) = e^{-\beta F(0)}$.
We allow the system to evolve from the initial time to some final time $t=T$. 
The forward evolution is governed by the evolution equation
\begin{eqnarray}
i \hbar \frac{dU_F(t,{\bf R})}{dt} = H(t, {\bf R}) U_F(t,{\bf R}).
\end{eqnarray}

If we consider the Heisenberg representation then the observables evolve 
as $A_F(t) = U_F(t)^{\dagger} A U_F(t)$ and similarly for the time-dependent 
Hamiltonian we have $H_F(t) = U_F(t)^{\dagger} H(t,{\bf R}) U_F(t)$.
During the backward process the external parameters are reversed. The system is
driven by a time-reversed Hamiltonian $H(T-t,-{\bf R})$ with a canonical 
density matrix 
\begin{eqnarray}
\rho(T) = \frac{e^{-\beta  H(T, -{\bf R}) }}{Z(T)}.
\end{eqnarray}
Note that the time dependence in the Hamiltonian is rescaled so that the 
initial time during the reverse process corresponds to $t=0$.
The partition function and the free energy for this process are given by 
the relation $Z(T)= {\rm Tr}[e^{-\beta  H(T, -{\bf R})}] = e^{-\beta F(T)}$.
The system is allowed to evolve until $t=T$ where the Hamiltonian at 
the end of the process is $H(0, -{\bf R})$. The evolution equation during the 
backward process is given by 
\begin{eqnarray}
i \hbar \frac{dU_R(t,{\bf R})}{dt} = H(T-t, {\bf R}) U_R(t,{\bf R}).
\end{eqnarray}
Now it can be shown that the forward and the backward evolution operators are
related via 
\begin{eqnarray}
\Theta U_F(T-t,{\bf R}) U_F^{\dagger}(t,{\bf R}) \Theta =  U_R(t, -{\bf R}).
\end{eqnarray}
In the Heisenberg picture the observables which have 
definite parity under time reversal obey the following relation 
during the forward and 
the backward process 
\begin{eqnarray}
A_F(t) = \pm   U_F^{\dagger}(T )\Theta A_R(T-t) \Theta  U_F(T),
\end{eqnarray}
where $A_R(T-t) = U_R^{\dagger}(T-t) A U_R(T-t)$.
Now, the universal work relation of Andrieux and Gaspard states the following.

\noindent
{\it Theorem}: For an arbitrary 
time-independent observable $A$ with a definite parity the functional relation 
\begin{eqnarray}
\langle e^{\int_0^T~dt \lambda(t) A_F(t) } e^{-\beta H_F(T)} 
e^{\beta H(0)} \rangle_{\rho(0),F} =
e^{-\beta \Delta F}  \langle e^{\pm \int_0^T~dt \lambda(T-t) A_R(t)} \rangle_{\rho(T),R}
\end{eqnarray}
connects the averages during the forward and the backward processes. Here, 
$\lambda(t)$ is an arbitrary function, $\Delta F = F(T) -  F(0)$ is the 
free energy difference of the two equilibrium states. The average of an 
operator $O$ during the
forward and the backward processes are defined via $<O>_{\rho(0),F} = 
{\rm Tr}[\rho(0) O]$ 
and $<O>_{\rho(T),R} =  {\rm Tr}[\rho(T) O]$, respectively.
As mentioned, the Jarzynski relation can be obtained from (7). 
This relation also provides the linear response theory of an arbitrary 
observable.

\section{ Inequality for quantum work relation}

In this section, we prove the main inequality for the universal work relation 
under the trial Hamiltonian.
Suppose that the system is not driven by the original Hamiltonian but by 
a different Hamiltonian $H'(t,{\bf R}')$ (we allow the new Hamiltonian to 
depend on a different set of external parameters ${\bf R}'$).
This can be a trial Hamiltonian or an approximate Hamiltonian or it could 
consist of perturbation to the original Hamiltonian. 
The question that we are concerned here is how stable or how accurate is the 
universal work relation with such a trial Hamiltonian that is 
different than the 
original one. To carry out the stability analysis of quantum work relation 
we use adiabatic representation theory where the original Hamiltonian 
$H(t,{\bf R})$ and the trial Hamiltonian $H'(t,{\bf R}')$ are  
assumed to have set of instantaneous eigenstates at the initial and the 
final time. Otherwise the results
presented here are quite general.  

Imagine that the system is driven by a trial Hamiltonian $H'(t,{\bf R}')$.
The system starts from a canonical density matrix 
\begin{eqnarray}
\rho'(0) = \frac{e^{-\beta  H'(0,{\bf R}') }}{Z'(0)}
\end{eqnarray}
at time $t=0$ and evolves as before until time $t=T$. During the 
forward process the 
observables evolve according to the trial Hamiltonian as 
$A_F'(t) = U_F'(t)^{\dagger} A U_F'(t)$ and similarly for the time-dependent 
Hamiltonian we have $H_F'(t) = U_F'(t)^{\dagger} H'(t,{\bf R}') U_F'(t)$. The 
backward process follows according to the trial Hamiltonian starting from a 
canonical density matrix 
\begin{eqnarray}
\rho'(T) = \frac{e^{-\beta  H'(T, -{\bf R}')}} {Z'(T)}.
\end{eqnarray}
Here, the primed quantities have similar meaning when the system is governed by
the trial Hamiltonian. The partition function is given by 
$Z'(T) = {\rm Tr}[e^{-\beta  H'(T, -{\bf R}')}]$. 
Now, with this trial Hamiltonian one can show that the 
universal work relation reads as

\begin{eqnarray}
\la e^{\int_0^T~dt \lambda(t) A_F'(t) } e^{-\beta H_F'(T)} e^{\beta H'(0)} 
\ra_{\rho'(0),F} =
e^{-\beta \Delta F'} \la e^{\pm \int_0^T~dt \lambda(T-t) A_R'(t)} \ra_{\rho'(T),R}
\end{eqnarray}
This connects the functional averages of an arbitrary observable 
during the forward and the backward 
process when the system is driven by a trial Hamiltonian. Here, 
$\lambda(t)$ being an arbitrary function, $\Delta F' = F'(T) - F'(0)$ being the 
free energy difference of the two equilibrium states. The averages during 
forward and the backward processes are defined as  
$\la O \ra_{\rho'(0),F} = {\rm Tr}[\rho'(0) O]$ 
and $\la O \ra_{\rho'(T),R} =  {\rm  Tr}[\rho'(T) O]$, respectively.

What we prove is that the following inequality is satisfied 

\begin{eqnarray}
\la e^{\int_0^T~dt \lambda(t) A_F'(t) } e^{-\beta H_F'(T)} e^{\beta H'(0)} 
\ra_{\rho'(0), F} \le 
e^{-\beta \Delta F}  e^{\beta (\la V(0)\ra_{\rho(0),F} - \la V(T)\ra_{\rho'(T),R} )} 
\la e^{\pm \int_0^T~dt \lambda(T-t) A_R'(t)} \ra_{\rho'(T), R}, \nonumber\\
\end{eqnarray}
where $V(t) = H'(t, {\bf R}') - H(t, {\bf R})$ is the difference between 
the trial and the actual Hamiltonian. The above inequality can also be 
stated as
$$
\frac{ \la e^{\int_0^T~dt \lambda(t) A_F'(t) } e^{-\beta H_F'(T)} e^{\beta H'(0)} 
\ra_{\rho'(0), F} } 
{\la e^{\pm \int_0^T~dt \lambda(T-t) A_R'(t)} \ra_{\rho'(T), R}}
\le 
\frac{ \la e^{\int_0^T~dt \lambda(t) A_F(t) } e^{-\beta H_F(T)} e^{\beta H(0)} 
\ra_{\rho(0), F} } 
{\la e^{\pm \int_0^T~dt \lambda(T-t) A_R(t)} \ra_{\rho(T), R}}
 e^{\beta (\la V(0)\ra_{\rho(0),F} - \la V(T)\ra_{\rho'(T),R} )}. 
$$

To prove the above inequality we assume that the time-dependent Hamiltonians 
$H(t)$ and $H'(t)$
have instantaneous eigenstates at time $t=0$ and $t=T$. 
Let the original Hamiltonian satisfies an eigenvalue equation at time $t=0$ as
$H(0, {\bf R}) |\psi_n(0) \ra = E_n(0,{\bf R}) |\psi_n(0) \ra $. The partition 
function at $t=0$ can be written as
$Z(0)= {\rm Tr} [e^{-\beta H(0, {\bf R})}] = 
\sum_n \langle \psi_n(0)| e^{-\beta H(0, {\bf R})} |\psi_n(0) \ra $. 
During the forward 
process consider the diagonal elements of the canonical density matrices 
$\rho(0)$ and $\rho'(0)$ in the eigenbasis $|\psi_n(0) \ra $. Thus, 
we have two probability distributions $p_n(0)$ and $p_n'(0)$ as given by
\begin{eqnarray}
p_n(0) &=& 
\frac{e^{-\beta E_n(0, {\bf R})}}{Z(0)} \nonumber\\
p_n'(0) &=& \la \psi_n(0)| \frac{e^{-\beta H'(0, {\bf R}')}}{Z'(0)} 
|\psi_n(0) \ra .
\end{eqnarray}

For these two probability distributions $p_n(0)$ and $p_n'(0)$ the relative 
entropy of $p$ with respect to $p'$ 
(also called the Kullback-Leibler distance) is defined by
\begin{eqnarray}
R(p, p') = \sum_n p_n(0) \log \frac{p_n(0)}{p_n'(0)} \ge 0.
\end{eqnarray}
It is a convex function of $p_n(0)$ and is always non-negative and 
equals zero only if $p_n(0) = p_n'(0)$. Using this, we can write
the inequality as
\begin{eqnarray}
\sum_n p_n(0) \log p_n(0) \ge \sum_n p_n(0) \log p_n'(0).
\end{eqnarray}

Now, using the expressions for $p_n(0)$ and $p_n'(0)$, we can write the above 
inequality as
\begin{eqnarray}
\sum_n p_n(0) [- \beta E_n(0,{\bf R}) - \log Z(0) ] \ge 
\sum_n p_n(0) \la \psi_n(0) |(- \beta H'(0,{\bf R}') - \log Z'(0) )
|\psi_n(0) \ra.
\end{eqnarray}
In the above equation, 
on the right hand side we have used the Jensen 
inequality, i.e., $\la e^O \ra \ge e^{\la O \ra}$ which gives us
\begin{eqnarray}
\log p_n'(0) &=& \log[ \la \psi_n(0) |(- \beta H'(0,{\bf R}') - 
\log Z'(0) )|\psi_n(0) \ra ]  \nonumber\\
&\ge & \la \psi_n(0) |(- \beta H'(0,{\bf R}') - 
\log Z'(0) )|\psi_n(0) \ra  
\end{eqnarray}
Therefore, we have the  following inequality that gives a lower bound on
the partition function with the trial Hamiltonian at the 
initial time of forward process 
\begin{eqnarray}
Z'(0) \ge Z(0) e^{-\beta {\rm Tr}[ \rho(0)~(H'(0, {\bf R}')  -  
H(0, {\bf R})  ] }.
\end{eqnarray}

Next we derive an upper bound on the partition function with the trial 
Hamiltonian at the initial time of the backward process. During the backward 
process with a trial Hamiltonian the system starts from the density matrix
as given in (9). Let us assume that the trial Hamiltonian satisfies an 
instantaneous eigenvalue equation at 
time $t=T$, i.e., $H'(T, -{\bf R}') |\psi_n'(T) \ra = E_n'(T, -{\bf R}') 
|\psi_n'(T) \ra$, where $E_n'(T, -{\bf R}')$ is the eigenvalue. 
Now consider the diagonal elements of the canonical density 
matrices corresponding to the Hamiltonian $H(T, -{\bf R})$ and 
$H'(T, -{\bf R}')$ in the eigenbasis of $|\psi_n'(T) \ra$. These probability 
distributions are given by
\begin{eqnarray}
p_n(T) &=& 
\la \psi_n'(T)| \frac{e^{-\beta H(T, -{\bf R}) }} {Z(T)} |\psi_n'(T)\ra 
 \nonumber\\
p_n'(T) &=& \la \psi_n'(T)| \frac{e^{-\beta H'(T, -{\bf R}')}}{Z'(T)} 
|\psi_n'(T) \ra 
= \frac{e^{-\beta E_n'(T, -{\bf R}')}}{Z'(T)}.
\end{eqnarray}

Let us consider the relative 
entropy of $p_n'(T)$ with respect to $p_n(T)$ which is given by 
\begin{eqnarray}
\sum_n p_n'(T) \log \frac{p_n'(T)}{p_n(T)} \ge 0.
\end{eqnarray}
Again using the non-negative property of the relative entropy 
we can write the above inequality as
\begin{eqnarray}
\sum_n p_n'(T) \log p_n'(T) \ge \sum_n p_n'(T) \log p_n(T).
\end{eqnarray}
Using the expressions for $p_n'(T)$, $p_n(T)$ and the Jensen inequality as
before, we have 
\begin{eqnarray}
\sum_n p_n'(T) [- \beta E_n'(T, -{\bf R}') - \log Z'(T) ] \ge \nonumber\\
\sum_n p_n'(T) \la \psi_n'(T) |(- \beta H(T, -{\bf R}) - 
\log Z(T) )|\psi_n'(T) \ra.
\end{eqnarray}
This leads to an upper bound for the partition function with the trial 
Hamiltonian during the time-reversed process as given by

\begin{eqnarray}
Z'(T) \le Z(T) e^{-\beta {\rm Tr}[\rho'(T) (~H'(T, -{\bf R}')  - 
 H(T, -{\bf R})~) ] }.
\end{eqnarray}

From these two inequalities (17) and (22) we have 
\begin{eqnarray}
\frac{Z'(T)}{Z'(0)} \le \frac{Z(T)}{Z(0)} e^{\beta [\la V(0) \ra_{\rho(0), F} - 
\la V(T) \ra_{\rho'(T), R}]},
\end{eqnarray}
where $\la V(0) \ra_{\rho(0),F} = {\rm Tr}[\rho(0)(~ H'(0, {\bf R}') - 
H(0, {\bf R})~) ]$ and 
$\la V(T) \ra_{\rho'(T),R} = {\rm Tr}[\rho'(T) (~H'(T, -{\bf R}') - 
 H(T, -{\bf R})~) ]$.  By noting the fact that 
$\frac{Z'(T)}{Z'(0)} = e^{ -\beta \Delta F'} $ and 
$\frac{Z(T)}{Z(0)} = e^{ -\beta \Delta F} $, we have the main inequality for 
the universal quantum work relation with a trial Hamiltonian as
given in Eq.(11). Hence, the proof.

\section{ Jarzynski relation with trial Hamiltonian}

In this section, we address the question of the accuracy of the Jarzynski 
relation when the system is driven away from equilibrium by a trial 
Hamiltonian.
Interestingly, the quantum version of the Jarzynski equality follows from the 
universal work relation of Andrieux and Gaspard \cite{gas}. If we set 
$\lambda=0$ in Eq.(7), we obtain
\begin{eqnarray}
 \la e^{-\beta H_F(T) } e^{\beta H(0) } \ra_F = \la e^{-\beta W} \ra 
= e^{-\beta \Delta F},
\end{eqnarray}
where the left hand side represents the average of the 
exponential of work performed
on the system during the forward process and $\Delta F$ is the equilibrium 
free energy difference $F(T)-F(0)$. Now if one asks the question what happens
to the Jarzynksi equality if we drive the system with a different Hamiltonian 
$H'(t)$ rather than $H(t)$. Here, $H'(t)$ could be a trial Hamiltonian, or
could be some perturbation. Under such a situation how stable is the Jarzynski 
relation or the nonequilibrium work relation. Our inequality can answer 
this question. It is clear that if we
drive the system with  a different Hamiltonian $H'(t, {\bf R}')$ then we
will have the Jarzynski equality 
\begin{eqnarray}
 \la e^{-\beta H_F'(T) } e^{\beta H'(0) } \ra_F = \la e^{-\beta W'} \ra= 
 e^{-\beta \Delta F'}.
\end{eqnarray}

From our inequality (11), if we set $\lambda =0$, we will obtain the following
inequality for the Jarzynski relation
\begin{eqnarray}
 \la e^{-\beta H_F'(T) } e^{\beta H'(0) } \ra_F = e^{-\beta \Delta F'}
\le e^{-\beta \Delta F} e^{\beta [\la V(0) \ra_{\rho(0),F} - 
\la V(T) \ra_{\rho'(T),R}]} \nonumber\\
{\rm or}~~~~ \la e^{-\beta W'} \ra \le \la e^{-\beta W} \ra 
e^{\beta [\la V(0) \ra_{\rho(0),F} -  \la V(T) \ra_{\rho'(T),R}]}.
\end{eqnarray}
This tells us that the work performed by a trial Hamiltonian in going 
from the initial configuration to the final configuration must 
respect this bound.
Eq.(26) is important in its own right. This shows that the correction term 
can be explicitly expressed as the averages of the difference operator 
between the accurate and trial Hamiltonians. By calculating the quantity 
$[\la V(0) \ra_{\rho(0),F} - \la V(T) \ra_{\rho'(T),R}]$ 
the error in the free energy difference caused 
by the modelling procedure can be estimated directly.

Furthermore, 
using the inequality (26), we can prove a generalized version of the 
Bogoliubov inequality. One version of the usual Bogoliubov inequality 
\cite{fey} can be stated as follows: Let 
$H$ be the original Hamiltonian and let $H'$ be the trial Hamiltonian with 
the condition that ${\rm tr}(\rho' H') = {\rm tr}(\rho' H)$, i.e., the averages 
of both the Hamiltonians in the canonical state $\rho'$ are same, then 
$F \le F'$. Here, $F$ is the free energy of the original Hamiltonian and 
$F'$ is the free energy of the trial Hamiltonian. The Bogoliubov inequality 
has important application in the mean field theory. For example, by a 
variational method if we can minimize the free energy of the trial Hamiltonian
we can get a better approximation to the exact free energy.

To prove the generalized version of the Bogoliubov inequality, 
we note from (26) that 
\begin{eqnarray}
 \Delta F' - \Delta F \ge [\la V(T) \ra_{\rho'(T),R} - 
\la V(0) \ra_{\rho(0),F}].
\end{eqnarray}
Now, if we impose the condition that $H(t)$ and $H'(t)$ have same averages 
in the original and trial canonical state at time $t=0$ and $t=T$, then 
$\la V(T) \ra_{\rho'(T),R} =0$ and $\la V(0) \ra_{\rho(0),F} =0$. Then 
it follows that 
\begin{eqnarray}
 \Delta F \le \Delta F'.
\end{eqnarray}
This can be regarded as time-dependent generalization of the 
Bogoliubov inequality which bounds the change in the free energy of 
original Hamiltonian with respect to perturbed one. 
This is an another important result. This inequality not only holds 
close to equilibrium but also hold in those situations which are 
far-from-equilibrium regime.

\section{Inequality with operator norm}

In this section, 
we prove an inequality for the Jarzynski relation which involves 
the norm of the operator $V(t)$. This may be useful when we do not have to 
calculate the averages of the operator $V(0)$ and $V(T)$ 
during the forward and backward 
processes, corresponding to the actual and trial Hamiltonians.

Consider the partition function for the trial 
Hamiltonian during the initial time of the reverse process. We have
\begin{eqnarray}
Z'(T) = {\rm Tr}[e^{-\beta  H'(T, -{\bf R}')} ]
\le {\rm Tr}[e^{-\beta  H(T, -{\bf R})} e^{-\beta  V(T)}],
\end{eqnarray}
where we have used the inequality ${\rm Tr}[e^{A + B} ] \le {\rm Tr}[e^{A} 
e^{  B} ]$ for all self-adjoint operators.

On expressing the trace using the eigenbasis of $H(T, -{\bf R})$ we have
\begin{eqnarray}
Z'(T) &\le & 
\sum_n \la \psi_n(T)|e^{-\beta  H(T, -{\bf R})}
  e^{-\beta  V(T)}| \psi_n(T) \ra \nonumber\\
&=&\sum_n \la \psi_n(T)|e^{-\beta  H(T, -{\bf R})}|\psi_n(T)\ra 
 \la \psi_n(T)| e^{-\beta  V(T)}| \psi_n(T) \ra.
\end{eqnarray}
Now, we use the inequality $||e^{A}|| \le e^{||A||}$ which holds for all 
operators. The operator norm of a linear operator $A$ is defined as
$|| A || = {\rm sup}_{(||\psi||=1)} || A(\psi) ||$. 	
(Note that in the matrix notation, $A(\psi) = A |\psi\rangle$ and $||A||$ is 
equal to the square root of the largest eigenvalue of the symmetric matrix 
$A^TA$ .)

Using this we have
\begin{eqnarray}
Z'(T) &\le&  
\sum_n \la \psi_n(T)|e^{-\beta  H(T, -{\bf R})} | \psi_n(T) \ra 
e^{\beta || V(T)|| }\nonumber\\
&=& Z(T) e^{\beta || V(T)||}.
\end{eqnarray}
This gives an upper bound for the partition function for the trial 
Hamiltonian at the initial time of the reverse process.

Now, let us consider the partition function $Z(0)$ 
with the Hamiltonian $H(0)$ during the initial time of the 
forward process. We have
\begin{eqnarray}
Z(0) = {\rm Tr}[e^{-\beta  H(0, {\bf R})} ]
\le {\rm Tr}[e^{-\beta  H'(0, {\bf R}')} e^{\beta  V(0)}],
\end{eqnarray}
where we have again used the inequality ${\rm Tr}[e^{A + B} ] \le {\rm Tr}[e^{A} 
e^{  B} ]$ which holds for all self-adjoint operators.
We express the trace using the eigenbasis of $H'(0, {\bf R})$. This gives us
\begin{eqnarray}
Z(0) & \le & 
\sum_n \la \psi_n'(0)|e^{-\beta  H'(0, {\bf R}')}
  e^{\beta  V(0)}| \psi_n'(0) \ra \nonumber\\
&=&\sum_n \la \psi_n'(0)|e^{-\beta  H'(0, {\bf R}')}|\psi_n'(0)\ra 
 \la \psi_n'(0)| e^{\beta  V(0)}| \psi_n'(0) \ra.
\end{eqnarray}
Now, we use the inequality $||e^{A}|| \le e^{||A||}$ which holds for all 
operators. Using this we have
\begin{eqnarray}
Z(0) &\le&  
\sum_n \la \psi_n'(0)|e^{-\beta  H'(0, {\bf R}')} | \psi_n'(0) \ra 
e^{\beta || V(0)|| }\nonumber\\
&=& Z'(0) e^{\beta || V(0)||}.
\end{eqnarray}
This gives a lower bound for the partition function with the trial Hamiltonian 
during the initial time of the forward process.

On multiplying two inequalities (31) and (34), we have the new 
inequality for the Jarzynski relation under the trial Hamiltonian as given by 
\begin{eqnarray}
\la e^{-\beta W'} \ra &=& e^{-\beta \Delta F'}  \le 
e^{-\beta \Delta F} e^{\beta || V(0)||} e^{\beta || V(T)||} \nonumber\\
{\rm or}~~~~
\la e^{-\beta W'} \ra & \le & \la e^{-\beta W} \ra 
e^{\beta || V(0)||} e^{\beta || V(T)||}.
\end{eqnarray}
This inequality may be often useful where we do not have to calculate 
explicitly the averages of the operator $V$ for the actual and trial 
states. 
Eq. (35) provides the exact estimation of the free energy difference between 
the initial and final states when the actual Hamiltonian is not the same as 
the trial or approximate Hamiltonian. With the help of the free energy, all 
the quantum equilibrium properties like different phases of a system and 
their physical properties like the phase transitions etc. can be exactly 
calculated. This relation is expressed  in terms of an inequality 
which involves a correction term when the actual Hamiltonian is not same 
as the trial or approximate Hamiltonian.  
The correction term is expressed in the form of the norm of the operator $V$
at the initial and final time. In some physical situations it may be difficult 
to calculate exactly the averages of $V(T)$ or $V(0)$. Also, the free energy
 of a quantum system relative to an arbitrary reference state is often 
difficult to determine. In such situations, Eq.(35)  may provide a 
practical tool to extract the quantum equilibrium 
information even if the averages of $V$ is explicitly not known. 
The norm of $V$ at the initial and final time is enough to give the 
required estimate.


\section{ Conclusions}

Testing various thermodynamic relations for complex systems are not 
always amenable.
One has to model the actual Hamiltonian of the system by a trial Hamiltonian 
and establish how accurate are those relations. In this paper, 
we have investigated the stability of the universal 
quantum work relation of Andrieux and Gaspard under a trial Hamiltonian 
and proved an inequality that must be respected if the system is driven away
from equilibrium by a trial Hamiltonian rather than the actual one. 
In turn, our inequality also tells us how accurate is the famous 
Jarzynski relation under such trial Hamiltonians. 
 We have shown, 
in particular, that 
the correction term can be explicitly expressed by the averages of the 
difference operator $V$ between 
the accurate and the trial Hamiltonians.
Therefore, the free energy difference caused by the modelling procedure 
can be estimated by calculating the difference operator. 
We have also obtained a generalized version of the 
Bogoliubov inequality for the time-dependent trial Hamiltonians. 
This tells us that 
the change in the free energy corresponding to the actual Hamiltonian is always 
less than or equal to the change in the free energy in the trial case.
Further, we have given an inequality using the operator norm where the 
correction term can be expressed in the form of the norm of the operator $V$
at the initial and final time.
In future, we plan to
apply this inequality to some realistic experiments. 
We hope that this inequality will be very useful in testing quantum work 
relations for many body systems and will open up further thoughts in the 
area of quantum work relations and fluctuation theorems. Also, this inequality 
may be applied to probe quantum nonequilibrium phenomena in glassy systems and 
many body systems.

\vskip 1cm

\noindent
{\it Acknowledgements:} AKP thank A. K. Rajagopal for useful 
discussions.



\end{document}